\documentclass{conm-p-l}
\usepackage{amsmath,amssymb,epsfig,latexsym}

\newcommand{\e}{\epsilon}

\newcommand{\ra}{\rightarrow}

\theoremstyle{definition}

\theoremstyle{remark}

\numberwithin{equation}{section}

\begin{document}
\title[Multiscale expansion of KdV]{
Numerical study of a multiscale expansion of KdV and Camassa-Holm equation}

\author{Tamara Grava}
\address{SISSA,  via Beirut 2-4, 34014 Trieste, Italy}
\email{grava@sissa.it}
\author{Christian Klein}
\address{Max Planck Institute for Mathematics in the 
Sciences, Inselstr. 22, 04103 Leipzig, Germany}
\email{klein@mis.mpg.de}
\thanks{We thank G.~Carlet, B.~Dubrovin and J.~Frauendiener for helpful 
discussions and hints. CK and TG acknowledge support by the MISGAM program 
of the European Science Foundation. TG  acknowledges support by the RTN ENIGMA and  Italian COFIN 2004 ``Geometric methods in 
the theory of nonlinear waves and their applications''.}

%    General info
\date{\today}

\dedicatory{This paper is dedicated to P.~Deift on the occasion of 
his 60th birthday.}


\begin{abstract}
    We study numerically solutions to the Korteweg-de Vries  and Camassa-Holm equation
    close to the breakup of the 
    corresponding solution   to the   dispersionless equation.
    The solutions are compared with the 
    properly rescaled   numerical solution to a fourth order ordinary differential 
    equation, the second member of the Painlev\'e I hierarchy. 
    It is shown that this solution gives a valid asymptotic 
    description of the solutions close to breakup. We present a detailed analysis of the 
    situation and compare the Korteweg-de Vries solution quantitatively with asymptotic 
    solutions obtained via the solution of the Hopf and the Whitham 
    equations. We give a qualitative analysis for the Camassa-Holm equation 
\end{abstract}
\maketitle

\section{Introduction}
It is well known that the solution of the Cauchy problem for the Hopf equation
\begin{equation}
\label{hopf}
u_t+6uu_x=0,\quad u(x,t=0)=u_0(x), \;\;x\in\mathbb{R},\;t\in\mathbb{R}^+
\end{equation}
reaches a point of gradient catastrophe in a finite time. The solution of the  viscosity or conservative 
regularization  of the above hyperbolic equation display a 
considerably different behavior.
Equation (\ref{hopf}) admits an Hamiltonian structure
\[
u_t +\{ u(x), H_0\}\equiv u_t +\partial_x\frac{\delta H_0}{\delta u(x)} =0,\quad 
\]
with Hamiltonian and Poisson bracket
\[
 H_0 =\int u^3\, dx, \quad
\{ u(x) , u(y)\}=\delta'(x-y),
\]
respectively.
All the Hamiltonian perturbations up to the order $\epsilon^4$ of the hyperbolic equation (\ref{hopf}) have been 
classified in \cite{dubcr}. They are parametrized by two arbitrary functions $c(u)$, $p(u)$ 
\begin{equation}
\begin{split}
\label{riem2}
&
u_t +6u\, u_x + \frac{\epsilon^2}{24} \left[ 2 c\, u_{xxx} + 4 c' u_x u_{xx}
+ c'' u_x^3\right]+\epsilon^4 \left[ 2 p\, u_{xxxxx} \right.\\
&
\\
&\left.
+2 p'( 5 u_{xx} u_{xxx} + 3 u_x u_{xxxx}) + p''( 7 u_x u_{xx}^2 + 6 u_x^2 u_{xxx} ) +2 p''' u_x^3 u_{xx}\right]=0,
\end{split}
\end{equation}
where the prime denotes the derivative with respect to $u$.
The corresponding Hamiltonian takes the form
\[
H=\int \left[ u^3 - \epsilon^2 \frac{c(u)}{24} u_x^2
+\epsilon^4  p(u) u_{xx}^2 \right]\, dx
\] 
For $c(u)=12$, $p(u)=0$ one obtains the Korteweg - de Vries (KdV) 
equation $u_t+6uu_x+\epsilon^{2}u_{xxx}=0$, and for $c(u)=48u $ and 
$p(u)=2u$  the Camassa-Holm equation up to order $\epsilon^{4}$;
for generic  choices of the functions $c(u)$, $p(u)$ equation (\ref{riem2}) 
is apparently not an integrable PDE. However it 
admits an infinite family of commuting Hamiltonians  up to order $O(\e^6).$

The case of small viscosity perturbations  
of one-component hyperbolic equations has been well studied and understood 
(see \cite{bressan} and references therein), while the  behavior of solutions to the conservative 
perturbation (\ref{riem2}) to the  best of our knowledge has not been investigated after the point of gradient catastrophe
of the unperturbed equation except for the  KdV case, 
\cite{LL,V2,DVZ}.

In a previous paper \cite{GK} (henceforth referred to as I) we have 
presented a quantitative numerical  comparison of 
the solution of the Cauchy problem for  KdV 
\begin{equation}
\label{KdV}
u_t+6uu_x+\epsilon^{2}u_{xxx}=0,\quad u(x,0)=u_0(x),
\end{equation}
in the small dispersion limit $\epsilon\ra 0$, and the 
asymptotic formula obtained in the works of Lax and Levermore \cite{LL}, Venakides \cite{V2} and 
Deift, Venakides and Zhou \cite{DVZ} which describes the solution of the 
above Cauchy problem at the leading order as $\e\rightarrow 0$. 
The asymptotic description of \cite{LL},\cite{DVZ}  gives in 
general a good approximation of the KdV solution, but is less  satisfactory near 
the point of gradient catastrophe of the hyperbolic  equation.
This problem has been addressed by Dubrovin in \cite{dubcr}, where, following  the universality 
results  obtained in the context of  matrix models by Deift et all \cite{DKMVZ}, he formulated the
universality conjecture about the behavior of a generic solution to
the Hamiltonian perturbation (\ref{riem2}) of the hyperbolic equation (\ref{hopf})
near the  point $(x_c,t_c,u_c)$ of gradient catastrophe for the 
solution of (\ref{hopf}).   
He argued that, up to shifts, Galilean transformations and 
rescalings, this behavior essentially depends neither on the choice 
of solution nor on the choice of the equation.
Moreover, the solution near the point $(x_c, t_c, u_c)$ is given by
\begin{equation}
\label{univer}
u(x,t,\e)\simeq u_c +a\,\epsilon^{2/7} U \left( b\, \epsilon^{-6/7} (x- x_c-6u_c (t-t_c)); c\, 
\epsilon^{-4/7} (t-t_c)\right) +O\left( \epsilon^{4/7}\right)
\end{equation}
where  $a$, $b$, $c$ are some constants that depend on the choice of the 
equation and the solution and  $U=U(X; T)$ is the unique real smooth  
solution to the fourth order ODE 
\begin{equation}\label{PI2}
X=6T\, U -\left[ U^3 + (\frac12 U_{X}^2 +  U\, U_{XX} ) +\frac1{10} U_{XXXX}\right],
\end{equation}
which is  the second member 
of the Painlev\'e I hierarchy. We will call this equation PI2.
The relevant solution  is characterized by the asymptotic behavior
     \begin{equation}
\label{PI2asym}
        U(X,T)=\mp(X)^{\frac{1}{3}}\mp \dfrac{2T}{X^{\frac13}}+O(X^{-1}),\quad X\ra \pm \infty,
\end{equation}
for each fixed $T\in\mathbb{R}$.
The existence of a smooth solution of (\ref{PI2}) for all  $X,T\in\mathbb{R}$ 
satisfying (\ref{PI2asym}) has been recently proved by Claeys and Vanlessen \cite{CV}.
Furthermore they study in \cite{CV1} the double scaling limit for the matrix model with the multicritical index and showed that the limiting  eigenvalues correlation kernel is obtained from the particular 
solution of (\ref{PI2}) satisfying  (\ref{PI2asym}). This result was conjectured in the 
work of Br\'ezin, Marinari and Parisi \cite{BMP}.

In this paper we address numerically the validity of (\ref{univer}) 
for the KdV equation, and we 
identify the region where this solution provides a  better 
description than the Lax-Levermore, and Deift-Venakides-Zhou theory. 
As an outlook for the  validity of (\ref{univer}) for other equations 
in the family (\ref{riem2}), we present a numerical analysis of the Camassa-Holm equation
near the  breakup point.
While the validity of  (\ref{univer})  can be theoretically proved using a Riemann-Hilbert approach to
the small dispersion limit of the KdV equation \cite{DVZ} 
and recent results in \cite{DKMVZ},\cite{CV},\cite{CV1}, for   the Camassa-Holm equation and also 
for the general Hamiltonian  perturbation to  the hyperbolic equation (\ref{hopf}), the problem is
completely open. Furthermore for the general equation (\ref{riem2}), the existence of a smooth solution 
for a  short time has not been established yet.
An equivalent analysis should also be performed  for Hamiltonian perturbation of elliptic systems, 
in particular for the semiclassical limit of the focusing nonlinear Schr\"odinger equation \cite{KMM},\cite{TVZ}.

The paper is organized as follows. In section 2 we give a brief 
summary  of  the   Lax-Levermore, and Deift-Venakides-Zhou theory 
and the  multiscale expansion (\ref{univer}). In section 3 we 
present the numerical comparison between the asymptotic description 
based on the Hopf and Whitham solutions and the multiscale solutions 
with the KdV solution. In section 4 we study the same situation for 
the Camassa-Holm equation. In the appendix we briefly outline the 
used numerical approaches.

\section{Asymptotic and multiscale solutions}
Following the work of \cite{LL}, \cite{V2} and  \cite{DVZ},
the rigorous theoretical  description of  the small dispersion limit of the 
KdV equation is the following:
Let $\bar{u}(x,t)$ be the zero dispersion limit of $u(x,t,\epsilon)$, namely
\begin{equation}
\label{baru}
\bar{u}(x,t)=\lim_{\e\ra 0}u(x,t,\e).
\end{equation}
\noindent
1) for $0\leq t< t_c$, where $t_c$ is a critical time,  
the  solution $u(x,t,\epsilon)$ of the KdV  Cauchy problem  is approximated, 
for small $\e$, by the limit $\bar{u}(x,t)$ which solves the Hopf equation 
\begin{equation}
\label{Hopf}
\bar{u}_t+6\bar{u}\bar{u}_x=0.
\end{equation}
Here $t_c$ is the time when the first
point  of gradient catastrophe appears in the solution 
\begin{equation}
\label{Hopfsol}
\bar{u}(x,t)=u_0(\xi),\quad x=6tu_0(\xi)+\xi,
\end{equation}
of the Hopf equation. 
From the above, the time $t_c$ of gradient catastrophe can be
evaluated from the relation
\[t_{c}=\dfrac{1}{\min_{\xi\in\mathbb{R}}[-6u_0'(\xi)]}.
\]
2) After the time of gradient catastrophe, 
the solution of the KdV equation is characterized by the
appearance  of an interval of rapid modulated oscillations.  
According to the Lax-Levermore theory, the interval $[x^-(t), x^+(t)]$ of the oscillatory zone is 
independent of $\epsilon$. Here $x^-(t)$ and $x^+(t)$  are
 determined from the initial data and satisfy the condition  $x^-(t_c)=x^+(t_c)=x_c$ where $x_c$ is the $x$-coordinate of the point of gradient catastrophe of the Hopf solution.
Outside the interval  $[x^-(t), x^+(t)]$ the leading order asymptotics of $u(x,t,\e)$  as $\e\ra 0$  is described by the solution of the Hopf equation (\ref{Hopfsol}).
Inside  the interval  $[x^-(t), x^+(t)]$ the solution $u(x,t,\e)$  is  approximately  
described, for small $\e$,  by  the elliptic solution of KdV 
\cite{GP}, \cite{LL}, \cite{V2}, \cite{DVZ},
\begin{equation}
\label{elliptic}
u(x,t,\e)\simeq \bar{u}+ 2\e^2\frac{\partial^2}{\partial
x^2}\log\theta\left(\dfrac{\sqrt{\beta_1-\beta_3}}{2\e K(s)}[x-2 t(\beta_1+\beta_2+\beta_3) -q];\mathcal{T}\right)
\end{equation}
where now $\bar{u}=\bar{u}(x,t)$ takes the form 
\begin{equation}
\label{ubar}
\bar{u}=\beta_1+\beta_2+\beta_3+2\alpha,
\end{equation}
\begin{equation}
\label{alpha}
\alpha=-\beta_{1}+(\beta_{1}-\beta_{3})\frac{E(s)}{K(s)},\;\;\mathcal{T}=i\dfrac{K'(s)}{K(s)},
\;\; s^{2}=\frac{\beta_{2}-\beta_{3}}{\beta_{1}-\beta_{3}}
\end{equation}
with  $K(s)$ and $E(s)$ the complete elliptic integrals of the first 
and second kind, $K'(s)=K(\sqrt{1-s^{2}})$;
 $\theta$ is the Jacobi elliptic theta function defined by the 
Fourier series
\[
\theta(z;\mathcal{T})=\sum_{n\in\mathbb{Z}}e^{\pi i n^2\mathcal{T}+2\pi i nz}.
\]
For constant values of the  $\beta_i$ the formula (\ref{elliptic}) is an exact solution of KdV well
known in the theory of finite gap integration \cite{IM}, \cite{DN0}. 
However  in  the description 
of the leading order asymptotics of $u(x,t,\e)$ as $\e\ra 0$,
 the quantities $\beta_i$ depend on $x$ and $t$  and evolve
 according to the Whitham equations \cite{W}
\[
\dfrac{\partial}{\partial t}\beta_i+v_i\dfrac{\partial}{\partial x}\beta_i=0,\quad i=1,2,3,
\]
where the speeds $v_i$ are given by the formula
\begin{equation}
    v_{i}=4\frac{\prod_{k\neq
     i}^{}(\beta_{i}-\beta_{k})}{\beta_{i}+\alpha}+2(\beta_1+\beta_{2}+\beta_{3}),
    \label{eq:la0}
\end{equation}
with $\alpha$ as in (\ref{alpha}). 
Lax and Levermore first derived, in the oscillatory zone, the expression (\ref{ubar}) for 
 $\bar{u}=\bar{u}(x,t)$ which clearly does not satisfy the Hopf equation. 
The theta function formula (\ref{elliptic}) for the leading order asymptotics 
of $u(x,t,\e)$ as $\e\ra 0$,  was obtained in the work of Venakides and the phase  
$q=q(\beta_1,\beta_2,\beta_3)$  was derived in the work of Deift, 
Venakides and Zhou \cite{DVZ}, using the steepest descent method for oscillatory Riemann-Hilbert problems \cite{DZh}
\begin{equation}
\label{q0}
    q(\beta_{1},\beta_{2},\beta_{3}) = \frac{1}{2\sqrt{2}\pi}
    \int_{-1}^{1}\int_{-1}^{1}d\mu d\nu \frac {f_-( \frac{1+\mu}{2}(\frac{1+\nu}{2}\beta_{1}
	+\frac{1-\nu}{2}\beta_{2})+\frac{1-\mu}{2}\beta_{3})}{\sqrt{1-\mu}
    \sqrt{1-\nu^{2}}},
\end{equation}
where $f_-(y)$ is the inverse function of the decreasing part of the initial data.
The above formula holds till some time $T>t_c$ (see \cite{DVZ} or  I for times 
$t>T$).

\noindent
3) Fei-Ran Tian proved  that the description  in 1) and 2) is generic 
for some time after  the time  $t_c$ of gradient catastrophe \cite{FRT1}.

\begin{figure}[!htb]
\centering
\epsfig{figure=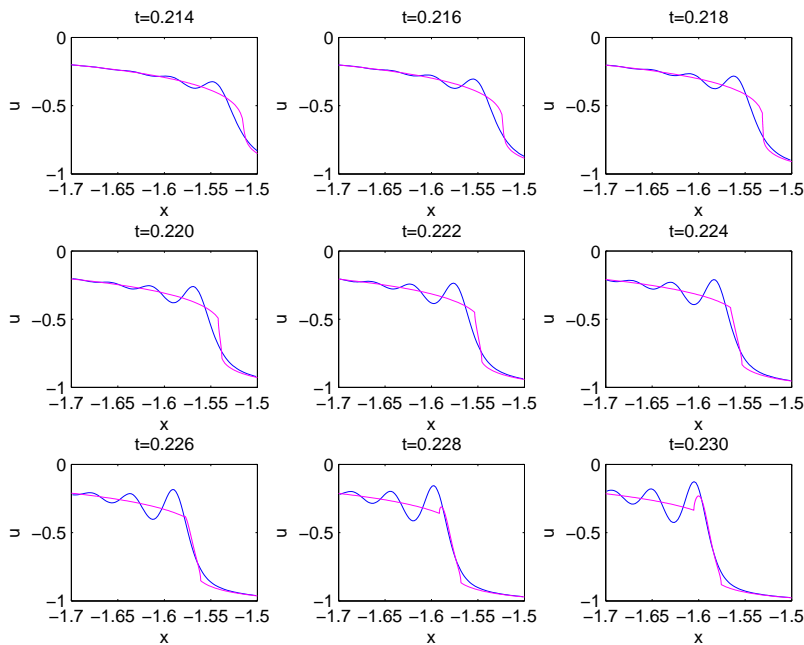, width=\textwidth}
\caption{The blue line is the solution of the KdV equation for the 
initial data $u_0(x)=-1/\cosh^2x$ and $\epsilon=10^{-2}$, 
and the purple line is the corresponding  leading order 
asymptotics given by formulas  (\ref{Hopfsol}) and (\ref{elliptic}).
The plots are given for different  times near the point of gradient catastrophe 
$(x_c,t_c)$ of  the Hopf solution. Here  $x_c\simeq -1.524 $, $t_c\simeq 0.216$.}
\label{fig2}
\end{figure}

\begin{figure}[!htb]
\centering
\epsfig{figure=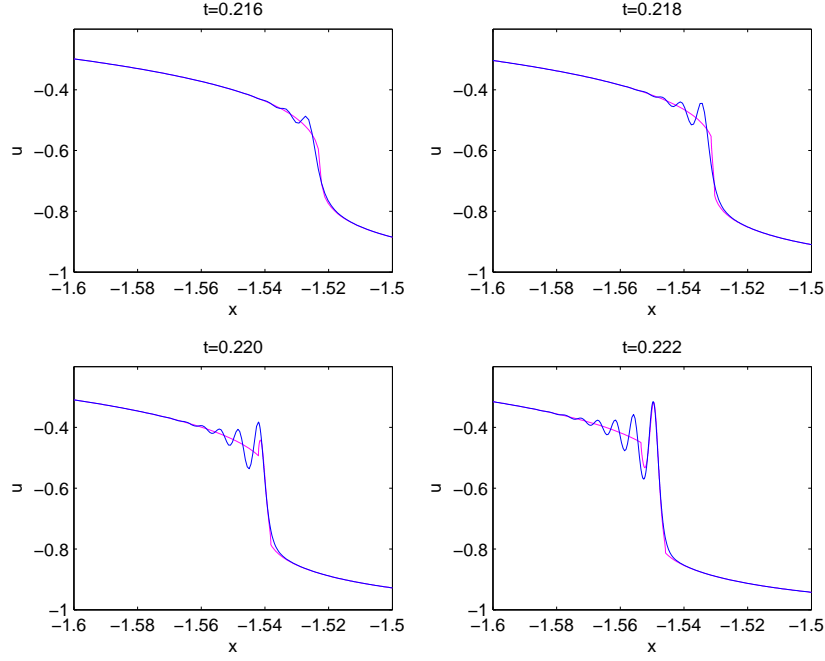, width=\textwidth}
\caption{KdV solution and asymptotic solution for $\epsilon=10^{-3}$ 
close to the breakup time.}
\label{figbreak4}
\end{figure}

In I we discussed the case $u_{0}(x) = -\mbox{sech}^{2}x$ in 
detail as an 
example. The main results were that the asymptotic description is of the order 
$\mathcal{O}(\epsilon)$ close to the center of the Whitham zone, but 
that the approach gives considerably less satisfactory results near the edges of the 
Whitham zone and close to the breakup of the corresponding solution 
to the Hopf equation. In the present paper we address the behavior 
near the point of gradient catastrophe of the Hopf solution in more 
detail. In Fig.~\ref{fig2} we show the KdV solution and the 
corresponding asymptotic solution as given above for several values 
of the time near the critical time $t_{c}$. It can be seen that there 
are oscillations before $t_{c}$, and that the solution in the Whitham 
zone provides only a crude approximation of the KdV solution for small 
$t-t_{c}$. 
The situation does not change in principle if we consider smaller 
values of $\epsilon$ as can be seen from Fig.~\ref{figbreak4}. The 
solution shows the same qualitative behavior as in Fig.~\ref{fig2}, 
just on smaller scales in $t$ and $x$.

\subsection{Multiscale expansion}
We give a brief summary of the results in \cite{dubcr} relevant for 
the KdV case we are interested in here. Near the 
point of gradient catastrophe $(x_c,t_c, u_c)$,
the
Hopf solution is generically given in lowest order by the cubic
\begin{equation}
\label{hopfsolc}
x-x_c \simeq 6(t-t_c) u -k(u-u_c)^3,\quad k=-f_-'''(u_c)/6,
\end{equation}
because $6t_c+f_-'(u_c)=0$ and $f_c''(u_c)=0$. Here $f_-(u)$ is the inverse of the 
decreasing part of the initial data $u_0(x)$. 
Now let us consider 
 $h_k=\dfrac{\delta H_k}{\delta u}$  where $H_k$ are the KdV  Hamiltonians 
such that $h_k=u^{k+2}/(k+2)!+O(\epsilon^2)$. We have
\[
h_{-1}=u,\;\;h_0=\dfrac{u^2}{2}+\dfrac{\epsilon^2}{6} u_{xx},\;\; h_1=\dfrac{1}{6}(u^3+\frac{\epsilon^2}{2}(u_x^2+2uu_{xx})+\dfrac{\epsilon^4}{10}u_{xxxx}),
\]
and the KdV equation is obtained from $u_t+6\partial_xh_0=0$.
Then
\[
x=6tu+a_0h_0+a_1h_1+\dots a_kh_k,
\]
is a symmetry of the KdV equation \cite{DZ}.
Setting $a_0=0,$ $a_1=-f'''(u_c)/6=k$ and $a_{k>2}=0$, and making the shift 
$t\rightarrow t-t_c$, $u\rightarrow u-u_c$ and the Galilean transformation  
$x\rightarrow x-x_c-6(t-t_c)u_c$  we arrive at the fourth order equation of Painlev\'e type
\begin{equation}
\label{painleve}
x-x_c-6(t-t_c)u_c = 6(t-t_c) (u-u_c) -k\left[(u-u_c)^3 +\epsilon^2
(\dfrac{ u_x^2}{2} +(u-u_c) u_{xx})+\dfrac{ \epsilon^4 }{10} u_{xxxx}\right]
\end{equation}
which is  an exact solution of the KdV equation and can be considered as a perturbation of the Hopf solution (\ref{hopfsolc})
near the point of gradient catastrophe $(x_c,t_c,u_c)$.
 The solution $u(x,t,\e)$ of (\ref{painleve}) is related to the solution $U(X,T)$ of (\ref{PI2}) 
by the rescalings
\begin{equation}
\label{KdVrescaled}
u(x,t,\e)=u_c+\left(\dfrac{\e}{  k}\right)^{2/7} U(X,T)
\end{equation}
where 
\begin{equation}
\label{scalings}
X =\dfrac{ x - x_c - 6 u_c (t-t_c)}{\epsilon^{\frac{6}{7}}k^{\frac17}},\quad 
T=\dfrac{t-t_c}{\e^{\frac{4}{7}}k^{\frac37}}.
\end{equation}
According to the  conjecture  in \cite{dubcr},  the solution (\ref{KdVrescaled}) is an approximation modulo terms  $O(\e^{\frac{4}{7}})$ to  the solution of the Cauchy problem (\ref{KdV}) for $(x,t,u)$ near the point of gradient catastrophe  $(x_c,t_c,u_c)$ of the hyperbolic equation (\ref{Hopf}).

\section{Numerical comparison}
In this section we will present a comparison of numerical solutions 
to the KdV equation and asymptotic solutions arising from solutions 
to the Hopf and the Whitham equations as well as the Painlev\'e I2 
equation as given above. Since we control the accuracy of the used 
numerical solutions, see I, \cite{numart1d} and the appendix, we 
ensure that the presented differences are entirely due to the 
analytical description and not due to numerical artifacts. We study 
the $\epsilon$-dependence of these differences by linear regression 
analysis. This will be done for nine values of $\epsilon$ between 
$10^{-1}$ and $10^{-3}$. Obviously the numerical results are only 
valid for this range of parameters, but it is interesting to note the 
high statistical correlation of the scalings we observe. 
We consider the initial data
\[
u_0(x)=-1/\cosh^2x.
\]
For this initial data 
\begin{equation}
     x_c=-\dfrac{\sqrt{3}}{2}+\log((\sqrt{3}-1)/\sqrt{2}),
\;\;t_c=\dfrac{\sqrt{3}}{8},\;\;
   \;\;u_c=-2/3.
    \label{tcrit}
\end{equation}

\subsection{Hopf solution}
We will first check whether the rescalings of the coordinates given 
in (\ref{KdVrescaled}) are consistent with the numerical results. It is 
known that the Hopf solution provides for times $t\ll t_{c}$ an 
asymptotic description of the KdV solution up to an error 
of the order $\epsilon^{2}$. This means that the $L_{\infty}$-norm of 
the difference between the two solutions decreases as $\epsilon^{2}$ 
for $\epsilon\ra 0$. For $t=0.1$ we actually observe this 
dependence. More precisely this difference $\Delta_{\infty}$
can be fitted  with a straight line by a standard linear 
regression analysis, 
$-\log_{10}\Delta_{\infty}=-a\log_{10}\epsilon+b$ with $a=1.9979$, 
with a correlation coefficient of $r=0.99999$ and standard 
error $\sigma_{a}=4.1*10^{-3}$. 

Near the critical time $t_{c}$ this picture is known to change 
considerably.  Dubrovin's conjecture \cite{dubcr} presented above 
suggests that the difference between Hopf and KdV solution near the 
critical point should scale roughly as $\epsilon^{2/7}$. 
In the following we will always compare solutions  in the intervals 
\begin{equation}
\label{interval}
[x_{c}+6u_{c}(t-t_{c})-\alpha 
\epsilon^{6/7},x_{c}+6u_{c}(t-t_{c})+\alpha \epsilon^{6/7}]
\end{equation}  where 
$\alpha$ is an $\epsilon$-independent constant (typically we take 
$\alpha=3$).

Numerically  we find at the critical time that the $L_{\infty}$-norm of the 
difference between Hopf and KdV solution scales like $\e^a$ where 
$a=0.2869$ ($2/7=0.2857\ldots$)  with correlation coefficient 
$r=0.9995$ and standard error $\sigma_{a}=6.9*10^{-3}$. 
Thus we confirm the 
expected scaling behavior within numerical accuracy. We also test 
this difference for times close to $t_{c}$. The relations 
(\ref{KdVrescaled}) suggest, however, a rescaling of the time, i.e., 
to compare solutions for different values of $\epsilon$ at the same 
value of $T$. We compute the respective solutions for KdV times 
$t_{\pm}(\epsilon)=t_{c}\pm 0.1\epsilon^{4/7}$. Before breakup at 
$t_{-}$ we obtain $a=0.31$ with $r=0.999$ and 
$\sigma_{a}=9.8*10^{-3}$, i.e., as expected a slightly larger value 
than $2/7$. After breakup at $t_{+}$ we find $a=0.26$ with $0.9995$ 
and $\sigma_{a}=6.6*10^{-3}$. We remark that after the breakup time,
the asymptotic solution is obtained by gluing the Hopf solution and the 
theta-functional solution (\ref{elliptic}).

\begin{figure}[!htb]
\centering
\epsfig{figure=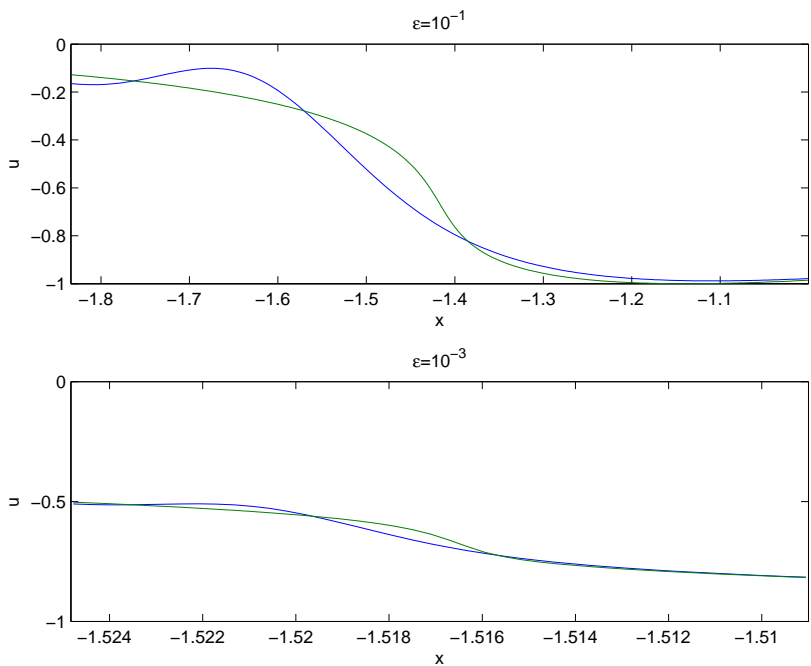, width=\textwidth}
\caption{KdV solution (blue) and Hopf solution (green) at the times 
$t_{-}(\epsilon)$ in a rescaled interval for two values of $\epsilon$.}
\label{figscalings}
\end{figure}

These results indicate that the scalings in (\ref{KdVrescaled}) are 
indeed observed by the KdV solution.  We show the corresponding situation for $t_{-}$ for two 
values of $\epsilon$ in Fig.~\ref{figscalings}.

\subsection{Multiscale solution}
\begin{figure}[!htb]
\centering
\epsfig{figure=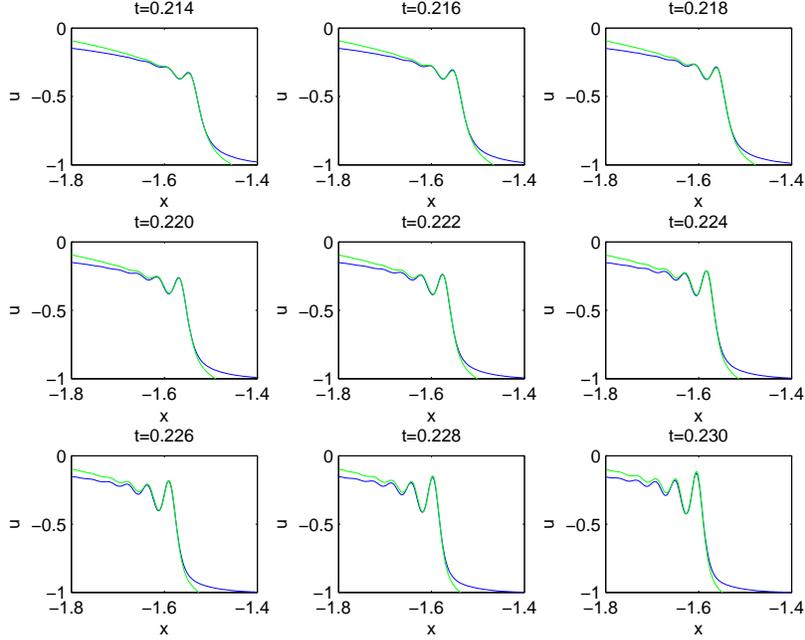, width=\textwidth}
\caption{The blue line is the solution of the KdV equation for the 
initial data $u_0(x)=-1/\cosh^2x$ and $\epsilon=10^{-2}$, 
and the green line is the corresponding  multiscale solution 
given by formula  (\ref{KdVrescaled}).
The plots are given for   different  times near the point of gradient catastrophe 
$(x_c,t_c)$ of  the Hopf solution. Here  $x_c\simeq -1.524 $, $t_c\simeq 0.216$.}
\label{figpain4breakg}
\end{figure}

\begin{figure}[!htb]
\centering
\epsfig{figure=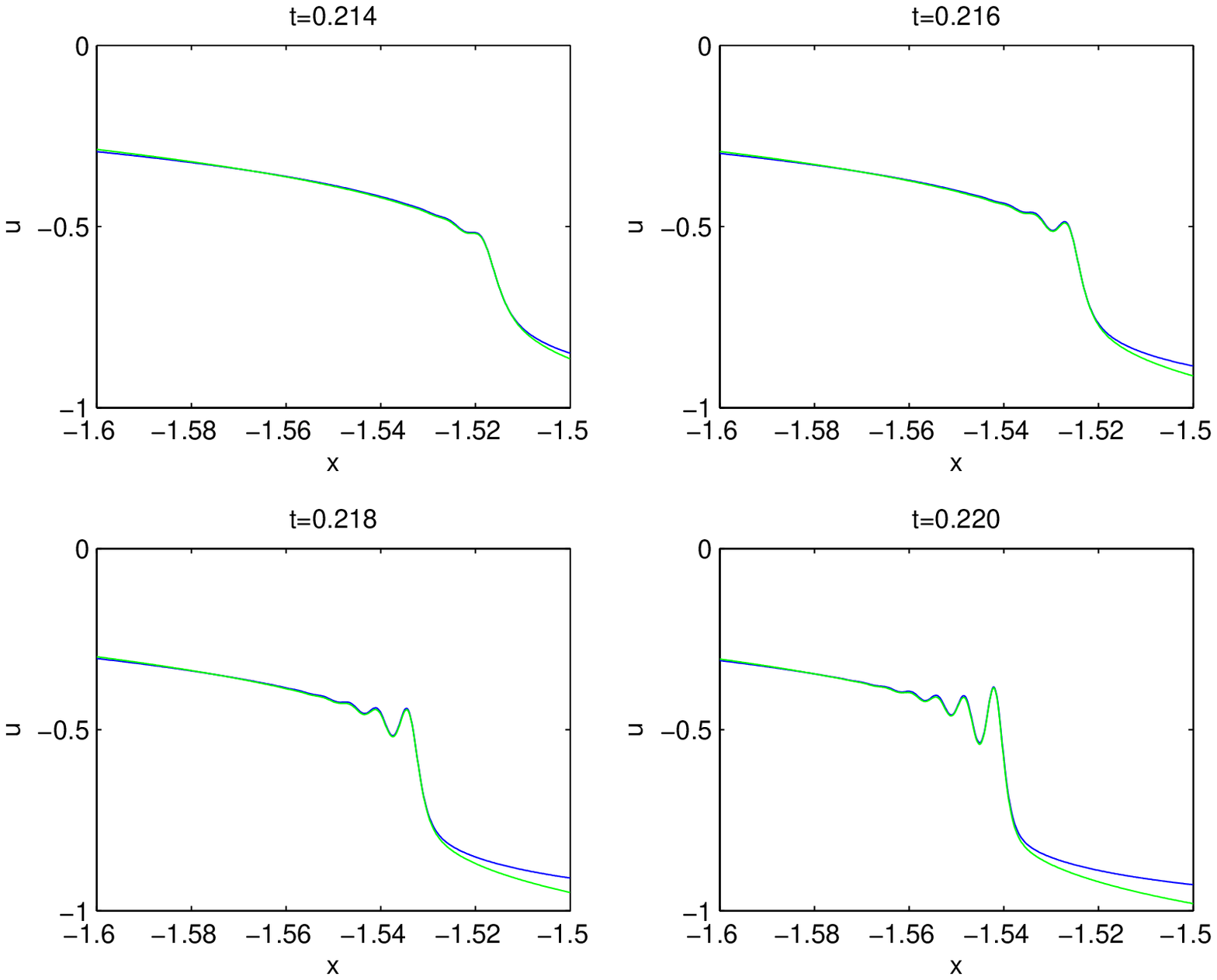, width=\textwidth}
\caption{The blue line is the solution of the KdV equation for the 
initial data $u_0(x)=-1/\cosh^2x$ and $\epsilon=10^{-3}$, 
and the green line is the corresponding  multiscale solution 
given by formula  (\ref{KdVrescaled}).
The plots are given for   different  times near the point of gradient catastrophe 
$(x_c,t_c)$ of  the Hopf solution.}
\label{figpain4break1e6}
\end{figure}

In Fig.~\ref{figpain4breakg} we show the numerical solution of the 
KdV equation for the initial data $u_{0}$ and the corresponding PI2
solution (\ref{KdVrescaled}) for $\epsilon=10^{-2}$ close to breakup. 
It can be seen that the PI2 solution (\ref{KdVrescaled}) gives a correct 
description of the KdV  solution close to the breakup point. 
For larger values of 
$|x-x_{c}|$ the multiscale solution is not a good approximation of the KdV solution.
A similar situation is shown in Fig.~\ref{figpain4break1e6} for the 
case $\epsilon=10^{-3}$. Obviously the approximation is better for 
smaller $\epsilon$. Notice that the asymptotic description is always better near 
the leading edge than near the trailing edge. 

\begin{figure}[!htb]
\centering
\epsfig{figure=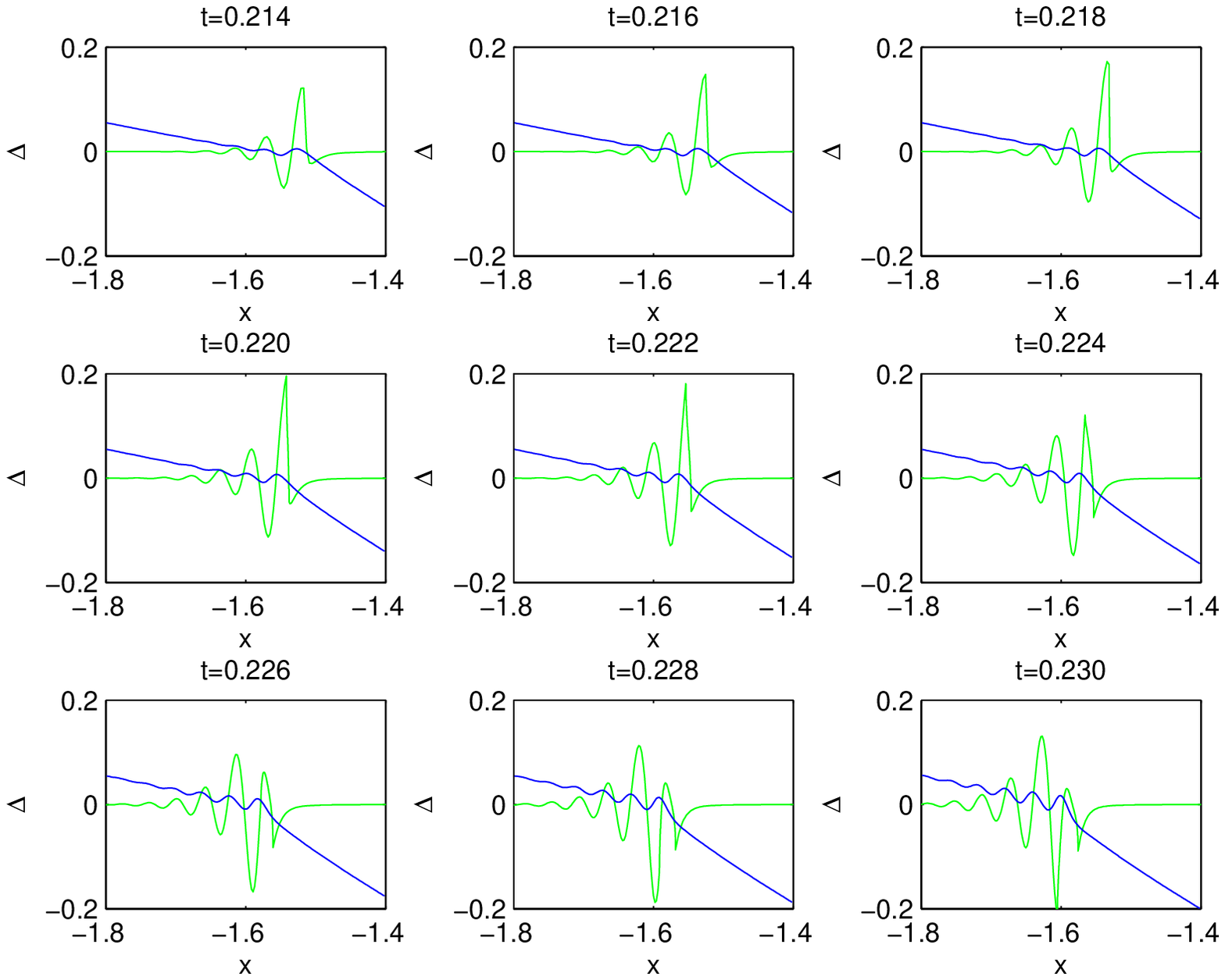, width=1.1\textwidth}
\caption{The blue line is the difference between the
solution of the KdV equation for the 
initial data $u_0(x)=-1/\cosh^2x$ and $\epsilon=10^{-2}$ and the 
multiscale solution, 
and the green line is the difference between the asymptotic solution 
and the KdV solution. 
The plots are given for different  times near the point of gradient catastrophe 
$(x_c,t_c)$ of  the Hopf solution. }
\label{figpain4delta}
\end{figure}
\begin{figure}[!htb]
\centering
\epsfig{figure=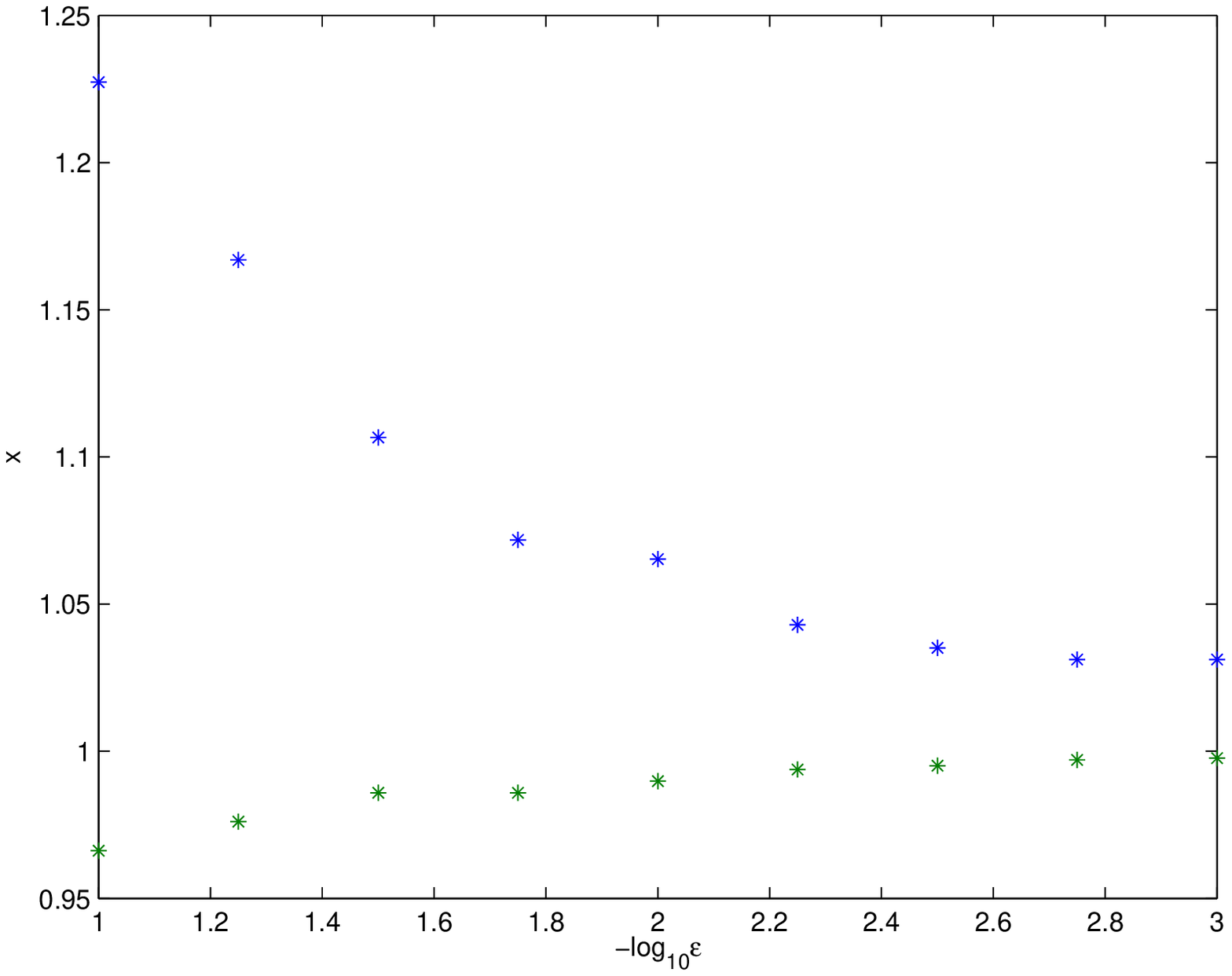, width=0.7\textwidth}
\caption{Limiting values of the zone where the multiscale solution 
provides a better asymptotic description of the KdV solution than the 
Hopf solution for $t=t_{c}$. The $x$ values are rescaled with $x_{c}$.}
\label{figwidthp12c}
\end{figure}
\begin{figure}[!htb]
\centering
\epsfig{figure=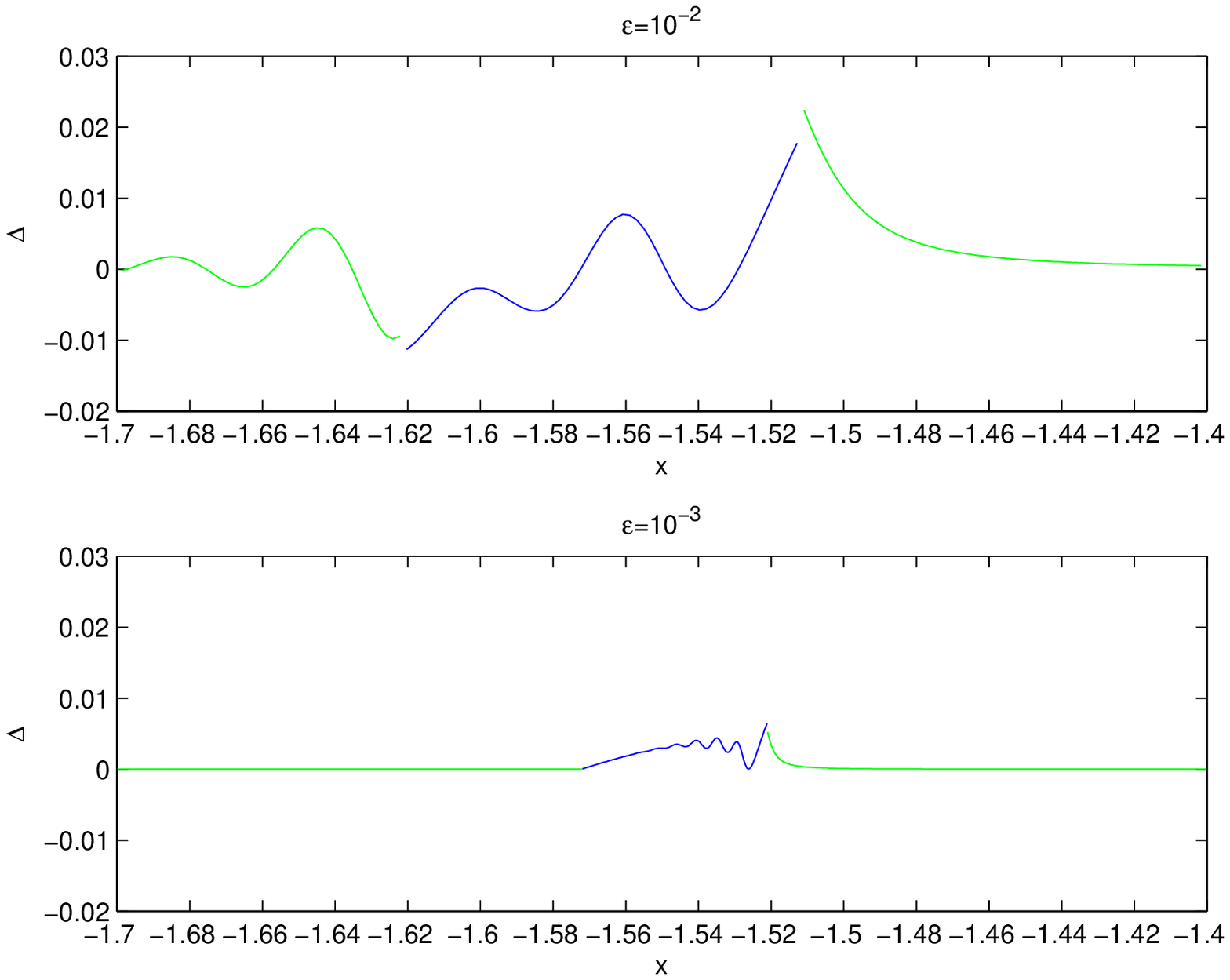, width=0.7\textwidth}
\caption{Difference of the KdV and the multiscale solution (blue) 
and the KdV and the Hopf solution (green)
for the initial data $u_0(x)=-1/\cosh^2x$ 
at $t=t_{c}$ for two values of $\epsilon$.}
\label{figdelta2ec}
\end{figure}

In  Fig.~\ref{figpain4delta} we plot in green
  the difference between the PI2 multiscale solution and the KdV 
solution  and in blue the difference between the KdV solution and  the asymptotic solutions 
(\ref{Hopfsol}) and (\ref{elliptic}).  It is thus possible to identify a zone around $x_{c}$ in which the 
multiscale solution gives a better asymptotic description. 
The limiting values of this zone rescaled by $x_{c}$ 
are shown in Fig.~\ref{figwidthp12c} for the critical time. 
It can be seen that the zone  always extends much further to the left (the direction of propagation) 
than to the right.
The width of this zone scales roughly as $\epsilon^{3/7}$, more 
precisely we find $\e^a$ with $a=0.468$, $r=0.981$ and $\sigma_{a}=0.073$.
We observe that the numerical scaling is smaller than the one predicted by the formula (\ref{scalings}). 
The matching of the multiscale and the Hopf solution can be seen in 
Fig.~\ref{figdelta2ec}. 

\begin{figure}[!htb]
\centering
\epsfig{figure=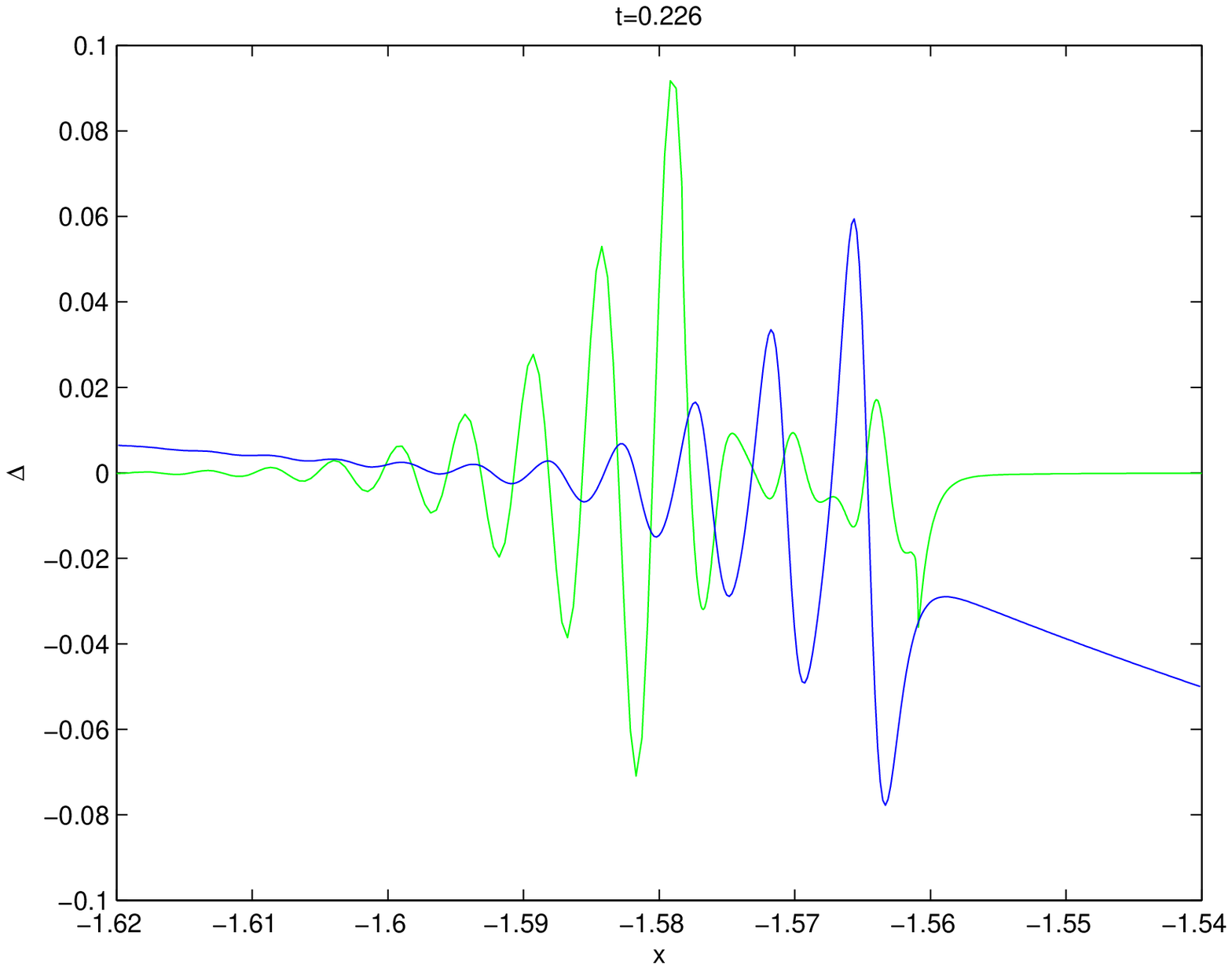, width=1.1\textwidth}
\caption{The blue line is the difference between 
solution of the KdV equation for the 
initial data $u_0(x)=-1/\cosh^2x$ and $\epsilon=10^{-3}$ and the 
multiscale solution, 
and the green line is the difference between the asymptotic solution 
and the KdV solution. 
The plots are given for $t=0.226$.}
\label{figpain4delta1e6.226}
\end{figure}

For larger times, the asymptotic solution (\ref{Hopfsol}) and (\ref{elliptic}) 
gives as expected a better  description of the KdV solution,  
see Fig.~\ref{figpain4delta1e6.226} for $\epsilon=10^{-3}$ and 
$t=0.226$. Close to the leading edge, the oscillations are, however, 
better approximated by the multiscale solution. 

To study the scaling of the difference between the KdV and the 
multiscale solution, we compute the $L_{\infty}$ norm of the 
difference between the solutions in the rescaled $x$-interval (\ref{interval}) with $\alpha=3$.
 We find that this error scales at the 
critical time roughly like $\epsilon^{5/7}$. More precisely we find 
a scaling  $\e^a$ where $a=0.708$ ($5/7=0.7143\ldots$) with correlation coefficient 
$r=0.9998$ and standard error  $\sigma_{a}=0.012$. 
Before breakup at the times $t_{-}(\epsilon)$ we obtain $a=0.748$ 
with $r=0.9996$ and $\sigma_{a}=0.016$, after breakup at the times 
$t_{+}(\epsilon)$ we get $a=0.712$ with $r=0.9999$ and 
$\sigma_{a}=6.2*10^{-3}$. Notice that the values for the scaling 
parameters are roughly  independent of the precise value of the 
constant $\alpha$ which defines the length of the interval (\ref{interval}).
 For instance  for $\alpha=2$, we find within the observed accuracy the same value.
In \cite{CV} Claeys and Vanlessen showed that the corrections to the 
multiscale solution appear in order $\epsilon^{3/7}$. For the values 
of $\epsilon$ we could study for our KdV example, the corrections are 
apparently of order $\epsilon^{5/7}$.

\section{Outlook}
The Camassa-Holm equation  \cite{CH} (see also
\cite{Fo})
\begin{equation}
\label{CH}
u_t+6uu_x-\epsilon^2(u_{xxt}+4u_xu_{xx}+2uu_{xxx})=0
\end{equation}
admits a bi-Hamiltonian description after the following Miura-type
transformation
\begin{equation}\label{cam-holm1}
m=u-\epsilon^2 u_{xx}.
\end{equation}
One of the Hamiltonian structure takes the form
\begin{equation}
\label{cam-holm-pb1}
\{ m(x), m(y)\} =\delta'(x-y) -\epsilon^2 \delta'''(x-y)
\end{equation}
so that the Camassa-Holm flow can be written in the form
\begin{equation}
\label{hamCH}
m_{t}=\{m(x), H\},\quad H= \int (u^3+uu_x^2)dx.
\end{equation}
To compare the Hamiltonian flow  in (\ref{riem2}) with the one given 
in (\ref{hamCH}) 
 one must first reduce the  Poisson bracket to the standard form $\{ \tilde u(x), \tilde u( y)\}_1=\delta'(x-y)$ by the transformation
\[
\tilde u =\left( 1-\epsilon^2 \partial_x^2\right)^{-1/2} m=m+\frac12 \epsilon^2 m_{xx} +\frac38 \epsilon^4 m_{xxxx}+\dots.
\]
After this transformation, the Camassa-Holm equation will take for 
terms up to order $\epsilon^{4}$ the form
\[
\tilde u_t +6\tilde u\, \tilde u_x +\epsilon^2 ( 8\tilde u_{x} \tilde u_{xx} + 4\tilde u\, \tilde u_{xxx}) +\epsilon^4( 20\, \tilde u_{xx} \tilde u_{xxx} + 12\, \tilde u_x \tilde u_{xxxx} +4\tilde u\, \tilde u_{xxxxx})+\dots=0 .
\]
which is equivalent to (\ref{riem2}) after the substitution
\[
c=48\tilde{u}, \quad p =2\tilde{u}.
\]
At the critical point $(x_c,t_c,u_c)$ the Camassa-Holm solution behaves according to the conjecture in 
\cite{dubcr} as
\[
u(x,t,\epsilon) = u_c -\left(\dfrac{\e^2|c_0|}{  k^2}\right)^{1/7} U(X,T)+O(\e^{\frac{4}{7}}),\quad c_0=4u_c
\]
where
\[
X =-\dfrac{1}{\epsilon}\left(\dfrac{\e}{k|c_0^3|}\right)^{1/7} 
( x - x_c - 6 u_c (t-t_c)),\quad
T=\left(\dfrac{1}{\e^4k^3c_0^2}\right)^{1/7}(t-t_c)
\]

\begin{figure}[!htb]
\centering
\epsfig{figure=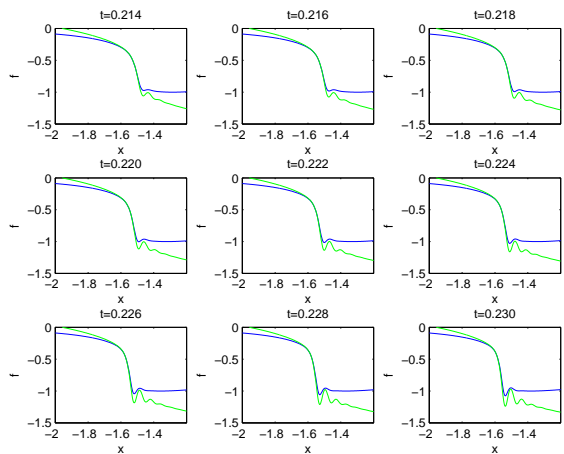, width=0.7\textwidth}
\caption{The blue line is the solution of the CH equation for the 
initial data $u_0(x)=-1/\cosh^2x$ and $\epsilon=10^{-2}$, 
and the green line is the corresponding  multiscale solution.
The plots are given for   different  times near the point of gradient catastrophe 
$(x_c,t_c)$ of  the Hopf solution. Here  $x_c\simeq -1.524 $, $t_c\simeq 0.216$.}
\label{figpainch9}
\end{figure}

In Fig.~\ref{figpainch9} we show the numerical solution to the CH 
equation for the initial data $u_{0}=-\mbox{sech}^{2}(x)$ and 
$\epsilon=10^{-2}$ at several values of time near the point of 
gradient catastrophe of the Hopf equation. It is interesting to 
compare this to the corresponding situation for the KdV equation in 
Fig.~\ref{figpain4breakg}. It can be seen that there are no 
oscillations of the CH equation on left side (the direction of the 
propagation) of the critical point, whereas in the KdV case all 
oscillations are on this side. The quality of the approximation of 
the CH and the KdV solution by the multiscale solution is also 
different. In the KdV case, the solution is well described by the 
multiscale solution on the leading part which includes the 
oscillations, whereas the approximation is less satisfactory on the 
trailing side. A similar behavior is observed in the CH case, but 
since the oscillations are now on the trailing side, they are not as 
well approximated as in the KdV case. The leading part of the 
solution near the critical point is, however, described in a better  
way.

\begin{figure}[!htb]
\centering
\epsfig{figure=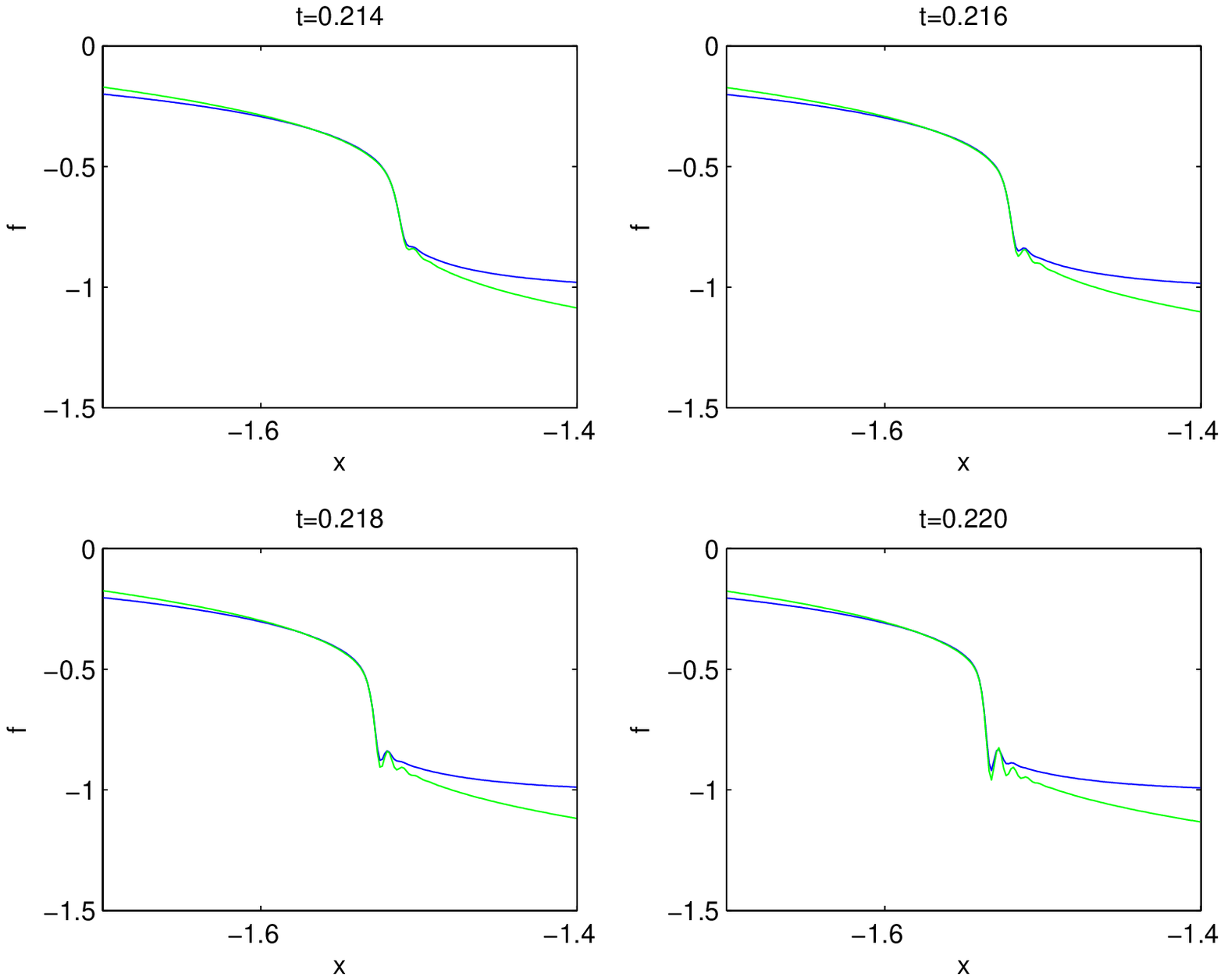, width=0.7\textwidth}
\caption{The blue line is the solution of the CH equation for the 
initial data $u_0(x)=-1/\cosh^2x$ and $\epsilon=10^{-3}$, 
and the green line is the corresponding  multiscale solution.
The plots are given for   different  times near the point of gradient catastrophe 
$(x_c,t_c)$ of  the Hopf solution.}
\label{figpainch4}
\end{figure}

The same qualitative behavior can also be observed for smaller 
$\epsilon$ in Fig.~\ref{figpainch4}, though the quality of the approximation increases as 
expected on the respective scales. Note that we plotted in 
Fig.~\ref{figpainch9} and Fig.~\ref{figpainch4} the CH solution 
instead of the function $\tilde{u}$, since there are no visible 
differences between the two for the used values of $\epsilon$.

\appendix
\section{Numerical solution of the fourth order equation}
We are interested in the numerical solution of the fourth order 
ordinary equation (ODE) (\ref{PI2})
with the asymptotic conditions (\ref{PI2asym}).
Numerically we will consider the equation on the finite interval 
$[X_{l},X_{r}]$, typically $X_{r}=-X_{l}=100$. In the exterior of this 
interval the solution to the equation (\ref{PI2}) is obtained in 
the form of a Laurent expansion of $F$ around infinity in terms of 
$Y=X^{1/3}$,
\begin{equation}
    U = Y + \sum_{n=1}^{\infty}\frac{(-1)^{n}a_{n}}{Y^{n}}.
    \label{pain3}
\end{equation}
We find the non-zero coefficients (not-given coefficients vanish)
$a_{1}= 2T,$ $a_{5}= - 8T^3/3$, $a_{6}= 1/18$,  $a_{7}=16T^4/3$, 
$a_{8}=- 5T/27$, $a_{10}=14T^2/27$, $a_{11}=- 256T^6/9$,  
$a_{12}=16T^3/3$, $a_{13}=640T^7/9-7/108$, \ldots This expansion also 
determines the boundary values we impose at $X_{l}$, $X_{r}$ for $U$ 
and $U_{X}$.

\begin{figure}[!htb]
\centering
\epsfig{figure=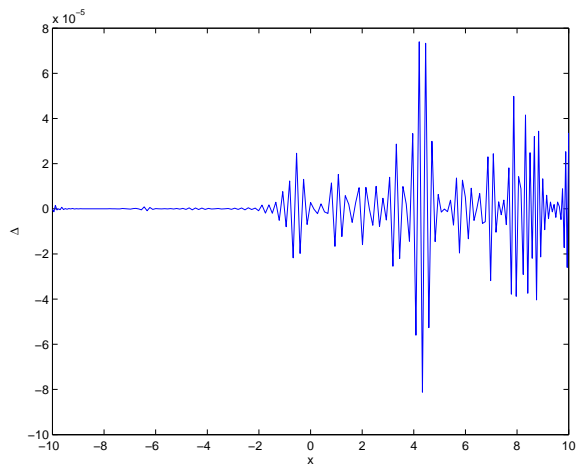, width=0.7\textwidth}
\caption{Residual of the numerical solution to the ODE (\ref{PI2}) 
for $t=0.23$. 
The derivatives are computed with Chebychev differentiation.} 
\label{figpain4test}
\end{figure}

The solution in the interval $[X_{l},X_{r}]$ is numerically obtained 
with a finite difference code based on a collocation method. The code 
\emph{bvp4c} distributed with Matlab, see \cite{bvp4c} for details, 
uses cubic polynomials in between the collocation points. The 
ODE (\ref{PI2}) is rewritten in the form of a first order 
system. With some 
initial guess (we use $U_{0}=-X^{1/3}$ as the initial guess), the 
differential equation is solved iteratively by linearization.  
The collocation points (we use up to 10000) 
are dynamically adjusted during 
the iteration. The iteration is stopped when the equation is 
satisfied at the collocation points with a prescribed relative accuracy, 
typically $10^{-6}$. The values of $U$ in between the collocation 
points are obtained via the cubic polynomials in terms of which the 
solution has been constructed. This interpolation leads to a loss in 
accuracy of roughly one order of magnitude with respect to the 
precision at the collocation points. To test this we determine 
the numerical solution via \emph{bvp4c} for (\ref{PI2}) 
on Chebychev collocation 
points and check the accuracy with which (\ref{PI2}) is satisfied 
via Chebychev 
differentiation, see e.g.\ \cite{trefethen1}. 
We are interested here in values of $|T|<1$ and $|X|<10$. It is found 
that the numerical solution with a relative tolerance of $10^{-6}$ on 
the collocation points satisfies the ODE to the order of better than 
$10^{-4}$, see Fig.~\ref{figpain4test} where we show the residual $\Delta$ by 
plugging the numerical solution into the differential equation. It is straight forward to 
obtain higher accuracy by requiring a lower value for the relative tolerance, 
but we will only need an accuracy of the solution of the order of 
$10^{-4}$ here.  
%%%%%%%%%%%
%%%%%%%%%%%%%%%%%
%%%%%%%%%%%%%%5

\end{document}